\documentclass[preprint2]{aastex}
\usepackage{amsmath}
\usepackage{natbib}
\usepackage{enumerate}
\usepackage{color}
\begin{document}

\title{A Significantly Low CO Abundance Toward the TW~Hya Protoplanetary Disk: A Path to Active Carbon Chemistry?}

\author{C\'ecile Favre,  L. Ilsedore Cleeves, Edwin A. Bergin,}
\affil{Department of Astronomy, University of Michigan, 500 Church St., 
    Ann Arbor, MI 48109}
\email{cfavre@umich.edu}
\author{Chunhua Qi,}
\affil{Harvard-Smithsonian Center for Astrophysics, 60 Garden Street, Cambridge, MA 02138}
\and
\author{Geoffrey A. Blake}
\affil{California Institute of Technology, Division of Geological \& Planetary
Sciences, MS 150-21, Pasadena, CA 91125, USA}
%
\begin{abstract}
In this Letter we report the CO abundance relative to H$_{2}$ derived toward the circumstellar disk of the T-Tauri star TW~Hya from the HD~($1-0$) and C$^{18}$O~($2-1$) emission lines. The HD~($1-0$) line was observed by the Herschel Space Observatory Photodetector Array Camera and Spectrometer whereas C$^{18}$O~($2-1$) observations were carried out with the Submillimeter Array at a spatial resolution of $2\farcs8 \times 1\farcs9$ (corresponding to $\sim142\times97$~AU). In the disk's warm molecular layer ($T>20$~K) we measure a disk-averaged gas-phase CO abundance relative to H$_2$ of $\chi{\rm(CO)}=(0.1-3)\times10^{-5}$, substantially lower than the canonical value of $\chi{\rm(CO)}=10^{-4}$. We infer that the best explanation of this low $\chi$(CO) is the chemical destruction of CO followed by rapid formation of carbon chains, or perhaps CO$_2$, that can subsequently freeze-out, resulting in the bulk mass of carbon locked up in ice grain mantles and oxygen in water.  As a consequence of this likely time-dependent carbon sink mechanism, CO may be an unreliable tracer of H$_2$ gas mass.
\end{abstract}


\maketitle
%
\section{Introduction}

Molecular hydrogen is the main gas-phase constituent in star-forming gas. However, it does not appreciably emit for typical gas conditions. Consequently carbon monoxide is widely used as a proxy for H$_{2}$ in the molecular interstellar medium \citep[e.g.][]{Dickman:1978} and protoplanetary disks \citep{Koerner:1995,Dutrey:1996}.
With a suite of transitions at millimeter/submillimeter wavelengths, the optically thick and thermalized $^{12}$CO lines trace gas temperature while optically thin CO isotopologues (namely $^{13}$CO and C$^{18}$O) probe the CO column and hence molecular mass.  A key component of the latter calculation is the calibration of CO to H$_{2}$, assuming an abundance of carbon monoxide, $\chi$(CO). In the ISM this factor can be constrained via comparisons of dust extinction to measurements of optically thin isotopologue lines. \citet{Ripple:2013} showed that typical $^{13}$CO abundances range from $\sim1-3\times10^{-6}$ in several clouds.  This corresponds to a CO abundance of $\sim0.6-2\times10^{-4}$, assuming an isotopic ratio of 60. 

Since the dense ISM provides CO to the protoplanetary disk during its formation, it is reasonable to assume that $\chi$(CO) in disks is similar to its interstellar value. Furthermore, at such high abundances, CO would represent the main gas-phase reservoir of carbon in disks.  Spatially resolved observations of CO could thus be used to determine the distribution and abundance of volatile carbon, which has implications for the inclusion of carbon into planetary systems \citep{lee10,Bond:2010,Oberg:2011}.

In this paper we combine spatially integrated observations of optically thin C$^{18}$O emission with a detection of the fundamental rotational transition of hydrogen deuteride, HD, towards the closest T-Tauri system, TW~Hya, at 51~pc \citep{Mamajek:2005}. HD emission provides a separate probe of H$_2$ \citep[][hereafter B13]{Bergin:2013}, with which we measure the $\chi$(CO) in this system.  We show that the main reservoir of {\em{gas-phase}} carbon, CO, is substantially reduced ($<10\%$ remaining) in the warm ($>20$~K) molecular layers of the disk and discuss implications of this result.

%
\section{Observations and data reduction}
\label{sec:obs}
The observations of TW~Hya were made on 2005 February 27 and April 10 using the Submillimeter Array\footnote{The Submillimeter Array is a joint project between the Smithsonian Astrophysical Observatory and the Academia Sinica Institute of Astronomy and Astrophysics and is funded by the Smithsonian Institution and the Academia Sinica.} \citep[SMA,][]{Ho:2004} located atop Mauna Kea, Hawaii. The SMA receivers operated in a double-sideband mode with an intermediate frequency (IF) band of  4--6~GHz from the local oscillator frequency, sent over fiber optic transmission lines to 24 overlapping ``chunks'' of the digital correlator. The correlator was configured to include CO, $^{13}$CO and C$^{18}$O, in one setting: the tuning was centered on the CO~($2-1$) line at 230.538~GHz in chunk S15, while the $^{13}$CO/C$^{18}$O~($2-1$) transitions at 220.399/219.560~GHz were simultaneously observed in chunks 12 and 22, respectively. CO~($2-1$) data were reported in \citet{Qi:2006}. Combinations of two array configurations (compact and extended) were used to obtain projected baselines ranging from 6 to 180~m. The observing loops used J1037--295 as the gain calibrator, with bandpass calibration using observations of 3C279. Flux calibration was done using observations of Titan and Callisto. 
Routine calibration tasks were performed using the MIR software package\footnote{http://www.cfa.harvard.edu/$\sim$cqi/mircook.html}, imaging and deconvolution were accomplished in the MIRIAD software package. The resulting synthesized beam sizes were $2\farcs8\times1\farcs9$ ($\rm{PA=-1.3\degr}$) and $2\farcs7\times1\farcs8$ (PA=-3.0$\degr$) for C$^{18}$O and $^{13}$CO, respectively. HD observations toward TW~Hya were been carried out with the Herschel Space Observatory Photodetector Array Camera and Spectrometer \citep{Poglitsch:2010,Pilbratt:2010}. Further informations concerning both reduction and line analysis are presented in B13.

\begin{figure}
\epsscale{0.9}
\plotone{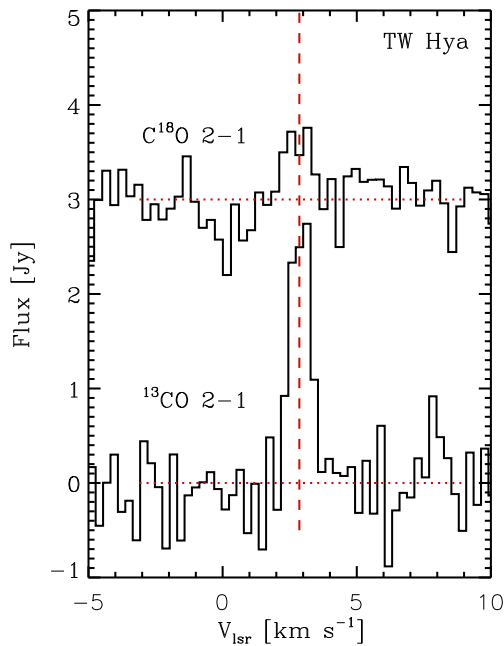}
\caption{Spatially integrated spectra of C$^{18}$O~($2-1$) (top) and $^{13}$CO~($2-1$) (bottom) in a 8$''$ square box centered on TW~Hya. The vertical dashed line indicates the LSR systemic velocity of the source (2.86~km~s$^{-1}$).\label{fig1}}
\end{figure}

In the present work we focus on the integrated line fluxes from HD, $^{13}$CO, and C$^{18}$O.  Spectroscopic parameters of these molecules and measured spectrally integrated fluxes within an 8$''$ box \citep[or 408~AU assuming a distance of 51~pc;][]{Mamajek:2005} are given in Table~\ref{tab1}.  
The spatially integrated spectra of C$\rm^{18}O$~($2-1$) and $^{13}$CO~($2-1$) are presented in Figure~\ref{fig1}.

\begin{table*}
\begin{center}
\caption{C$^{18}$O, $^{13}$CO and HD spectroscopic line parameters\tablenotemark{a} and total integrated fluxes observed towards TW~Hya. \label{tab1}}
\begin{tabular}{llllll}
\tableline\tableline
Molecule\tablenotemark{a} &Frequency&Transition&A&E$_u$&F\tablenotemark{b}\\
& (GHz) & &  (10$^{-8}$ s$^{-1}$) & (K) &  (10$^{-18}$ W~m$^{-2}$)\\
\tableline
HD & 2674.986& 1--0  & 5.44 &  128.38 & (6.3$\pm$0.7)\tablenotemark{c} \\
C$^{18}$O & 219.560 & 2--1 & 60.12 & 15.81  & (6.0$\pm$1.3)$\times$10$^{-3}$\\ 
$^{13}$CO & 220.399 & 2--1 & 60.74  & 15.87  &(20.0$\pm$1.3)$\times$10$^{-3}$\\
\tableline
\tablenotetext{a}{All spectroscopic data from $^{13}$CO, C$^{18}$O and HD are available from the CDMS molecular line catalog \citep{Muller:2005} through the Splatalogue portal \cite[www.splatalogue.net,][]{Remijan:2007} and are based on laboratory measurements and model predictions by \citet{Goorvitch:1994,Klapper:2000,Klapper:2001,Cazzoli:2004,Pachucki:2008,Drouin:2011}.}
\tablenotetext{b}{The total integrated fluxes are given with 1$\sigma$ uncertainty, which includes the calibration uncertainty.}
\tablenotetext{c}{From B13.}
\end{tabular}
\end{center}
\end{table*}

%
\section{Analysis}
\label{sec:analysis}
In the present study, we derive TW~Hya's disk-averaged gas-phase CO abundance from the observed C$^{18}$O~($2-1$) and HD~($1-0$) lines.  The conversion from integrated line intensity to physical column density is dependent on optical depth and temperature. In the following sections we explore a range of physically motivated parameter space assuming the emission is co-spatial and in LTE.  Based upon these assumptions we calculate a range of $\chi$(CO) in the warm (T~$>20$~K) disk using HD as our gas mass tracer.  Caveats of this approach and their implications for our measurement will be discussed in Section~\ref{caveats}.

\subsection{Line Opacity}
\label{sec:opa}
The determination of the CO mass from the C$^{18}$O emission relies on the assumption that C$^{18}$O~($2-1$) is optically thin and an $\rm{^{16}O/^{18}O}$ ratio. To estimate the disk-averaged opacity of C$^{18}$O, we compare C$^{18}$O~($2-1$) to $^{13}$CO~($2-1$) and find the disk-averaged $^{13}$CO/C$^{18}$O flux ratio is $\sim$3.3 $\pm$ 0.9. This measurement is strongly affected by the opacity of $^{13}$CO~($2-1$), where $\rm{\tau(^{13}CO)\sim2.9}$ assuming isotope ratios of $\rm{^{12}C/^{13}C=70}$ and $\rm{^{16}O/^{18}O=557}$ for the local ISM \citep{Wilson:1999}. This ratio suggests that the spatially integrated C$\rm^{18}O$ emission is thin, $\rm{\tau(C^{18}O)\sim0.36}$. 

\subsection{Hints from Disk Models}
The mismatch between the normal CO abundance and mass needed to match HD can be understood by computing the optically thin C$^{18}$O emission predicted by the sophisticated \citet{Gorti:2011} model.
For this purpose we adopt the non-LTE code {\it{LIME}} \citep{Brinch:2010} with the \citet{Gorti:2011} physical structure employed in the original modeling effort of \citet{Bergin:2013}, which best matched the HD emission.  In these calculations we include CO freeze-out assuming a binding energy of 855~K \citep{Oberg:2005}. The disk model natively assumes $\chi$(CO)$=2.5\times10^{-4}$ and if one adopts ${\rm^{16}O/^{18}O=500}$, over-predicts the C$^{18}$O~($2-1$) flux by $\sim$10$\times$.  Furthermore, the C$^{18}$O~($2-1$) emission is predicted to be optically thick and, to match the observed flux, $\chi$(C$^{18}$O) needs to be reduced to $\sim7\times10^{-9}$, i.e., $\rm{\chi(CO)=4\times10^{-6}}$.  We note that this abundance is dependent on the assumed binding energy, discussed further in Section~\ref{sec:freeze}.


\subsection{Mass and Model Independent $\chi$(CO) Determination}
\label{sec:coabund}
Under the assumption of optically thin  HD~($1-0$) and C$^{18}$O~($2-1$) emission, we can define the observable $R_{\rm{obs}}$ as the ratio between the {\em{observed}} number (denoted ${\mathcal{N}}$) of C$^{18}$O and HD molecules in their respective upper states, 
\begin{equation}
R_{\rm{obs}}=\frac{{\mathcal{N}}({\rm{C^{18}O}},J_u=2)}{{\mathcal{N}}({\rm{HD}},J_u=1)}=\frac{F_{\rm{C^{18}O}}A_{\rm{HD}}\nu_{\rm{HD}}}{F_{\rm{HD}}A_{\rm{C^{18}O}}\nu_{\rm{C^{18}O}}},\label{eq:ratio1}\end{equation}
where $\nu\rm_{X}$, $A\rm_{X}$ and $F\rm_{X}$ are the frequency, Einstein A coefficient and total integrated flux of the measured transition, respectively (see Table~\ref{tab1}).  To determine the total CO abundance in LTE, we must calculate the fractional population in the upper state, $f_u({\rm X})$, and assume isotopic ratios.  We adopt the isotopic oxygen ratio described in Section~\ref{sec:opa} and an isotopic ratio of HD relative to H$_2$ of $\chi({\rm{HD}})=3\times10^{-5}$, based on a D/H elemental abundance of ($1.50\pm0.10)\times10^{-5}$ \citep{Linsly:1998}. Assuming LTE and inserting the measured fluxes, the $^{12}$CO abundance relative to H$_2$ can be written as:
\begin{align}
\chi({\rm{CO}})&=1.76\times10^{-5}\left(\frac{{\rm^{16}O/^{18}O}}{557}\right)\left(\frac{R_{\rm{obs}}}{1.05\times10^{-3}}\right)\nonumber\\\times&\left(\frac{\chi({\rm{HD}})}{3\times10^{-5}}\right)\frac{f_u({\rm{HD}},J_u=1)}{f_{u}({{\rm{C^{18}O}}},J_u=2)}.\label{eq:abu}
\end{align}

It is important to note that the above analysis hinges upon the assumption that HD~($1-0$) and C$^{18}$O~($2-1$) are in LTE. Based on the \citet{Gorti:2011} model, at radii between $R\sim50-150$ AU the typical H$_2$ density at gas temperatures near $T_g=30$ K ranges between $\sim10^6-10^8$~cm$^{-3}$. Critical densities for the HD~($1-0$) and C$^{18}$O~($2-1$) transitions are respectively $2.7\times10^3$~cm$^{-3}$ and $10^4$~cm$^{-3}$ at $T_g\sim30$ K, respectively, which assumes collision rate coefficients with H$_2$ at 30~K of HD \citep[$2\times10^{-11}$~cm$^3$~s$^{-1}$;][]{Flower:2000} and C$^{18}$O \citep[$6\times10^{-11}$~cm$^3$~s$^{-1}$;][]{yang10}.
Under these conditions both lines are thermalized and the assumption of LTE is reasonable. 

The measured gas-phase disk-averaged $\chi({\rm CO})$, Eq.~\ref{eq:abu}, depends sensitively on the temperature of the emitting material, viz., the upper state fraction, $f_u$.  Formally, gas temperatures vary by orders of magnitude throughout the disk. However, to first order, as a result of the abundance distribution and excitation of a given rotational transition, emission generally arises from a narrower range of temperatures.  There are two ways temperatures can be estimated: 1) by observing optically thick lines originating from the same gas and measuring an average kinetic temperature of the emitting gas within the beam, or 2) by inferring temperatures from disk thermochemical models.

\begin{figure*}\epsscale{2}\plotone{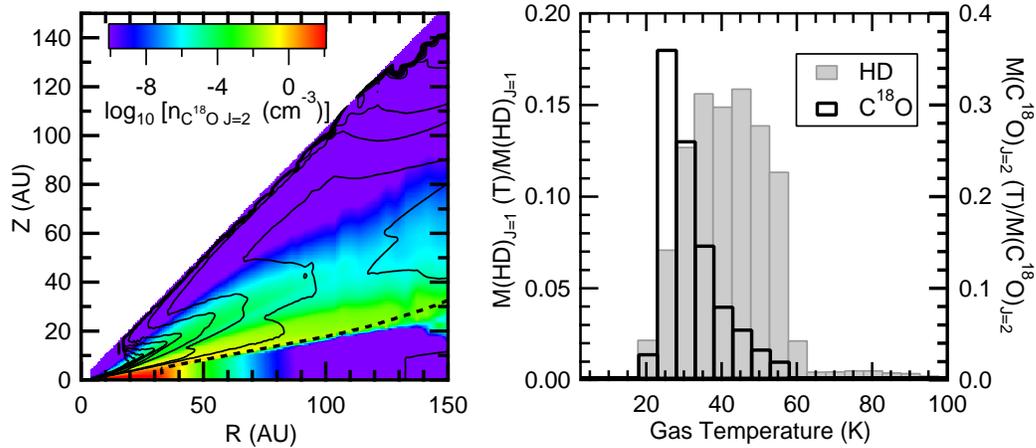}
\caption{{\em Left:}  Image details the radial and vertical distribution of the C$^{18}$O~$J=2$ volume density, $n_{\rm C^{18}O\;J=2}$, predicted in the \citet{Gorti:2011} model with $M_{\rm{gas}}=0.06~M_{\odot}$ and $\chi({\rm{C^{18}O}})=7\times10^{-9}$.  Contours are the gas temperature structure at 10, 20 (dashed line), 50, 75, 100, 150, 200, 250, and 300 K, respectively.
{\em Right:} Mass fraction of HD (gray) and C$^{18}$O (unfilled) in their respective upper states arising from gas at the specified temperature from the non-LTE calculation.  The mass is normalized to the total mass in the upper state $J_u$. For further details see Section~\ref{sec:coabund}}.\label{fig2}\end{figure*}

In the latter case we estimate a characteristic temperature of CO by dividing up the {\it{emissive}} mass of the \citet{Gorti:2011} model into temperature bins for both C$^{18}$O and HD, Figure~\ref{fig2}.
%
To compute the emissive mass we: following the \citet{Gorti:2011} TW~Hya model, for each temperature bin integrate the mass in HD $(J=1)$ and in C$^{18}$O $(J=2)$ {\it{in the upper state}} within the specified temperature range, and normalize this to the total mass throughout the disk in the corresponding upper state for each species, i.e., $M_{\rm{upper}}({\rm HD})=4\pi\int{n_{{\rm{HD}}\;J=1}rdrdz}$, where $n_{\rm{HD}\;J=1}$ is the upper state volume density calculated from the LIME excitation models \citep{Brinch:2010} performed for HD in B13.  

One notable feature of Figure~\ref{fig2} is that the two lines have slightly different peak maximally emissive temperatures, $\sim 20$~K for C$^{18}$O and $\sim 40-60$~K HD. However, over the temperature range expected for HD, the difference in the ratio of fractional populations for HD and C$^{18}$O is not enough to bring the CO abundance close to 10$^{-4}$ using Equation~\ref{eq:abu}.

\begin{figure}\epsscale{1.0}\plotone{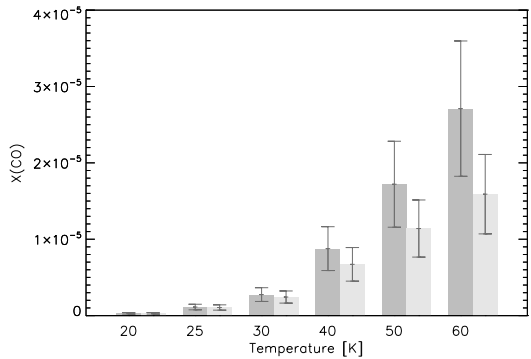}
\caption{CO Abundance with respect to H$_2$ as a function of emitting temperature within the warm molecular layer.  Light gray histograms show $\chi$(CO) for fixed ${\rm{T_{ex}(C^{18}O)=20}}$~K and T$_{\rm{ex}}$(HD) in the range 20--60K. Dark gray histograms show $\chi$(CO) for  ${\rm{T_{ex}(C^{18}O)=T_{ex}(HD)}}$. One sigma error bars taking into account the calibration uncertainty are shown.\label{fig3}}
\end{figure}

Guided by this range of temperatures, we compute the $\chi$(CO) from Eq.~\ref{eq:abu} assuming a C$^{18}$O gas temperature of $T_g=20$~K and varying the HD emitting temperature $T_g$(HD) between 20 and 60~K, accounting for the possibility of HD emitting from warmer gas than the C$^{18}$O.  The obtained $\chi$(CO) is provided in Fig.~\ref{fig3}. In all cases, $\chi\rm(CO)$ in the gas is lower than the canonical value of $\chi\rm(CO)\sim10^{-4}$; ranging between $(0.1-3)\times10^{-5}$.
From the modeled mass distribution shown in Fig.~\ref{fig2}, the center of the gas temperature distribution probed by HD is $T_g\sim40$~K, while  C$^{18}$O mostly emits from 20~K. With this value the resulting CO abundance is only ${\rm{\chi(CO)=7\times10^{-6}}}$, over 10$\times$ less than the canonical value. 

Therefore, to get $\chi$(CO) up to the 10$^{-4}$ range, significant corrections to the upper state fraction of each species is required. Concerning C$^{18}$O, that requires the gas to be either significantly colder or hotter such that the $J=2$ becomes depopulated. Both scenarios are unlikely (see Fig.~\ref{fig2}) and unsupported by the $^{12}$CO data \citep{Qi:2006}.

Alternatively, $^{12}$CO emission can constrain the temperature in the layers where its emission becomes optically thick.
Using the resolved Band 6 TW~Hya ALMA Science Verification (S.V.) observations of $^{12}$CO~($2-1$), the peak beam temperature is 24.5~K within a 2$\farcs83\times2\farcs$39 ($\rm{P.A.}=44\degr$) beam. This temperature represents the beam averaged kinetic temperature of the CO emitting gas within the inner $R\sim70$~AU, in agreement with values reported by B13 for the Band 7 S.V. data of the CO ($3-2$) line and the observations of \citet{Qi:2006} for CO~($6-5$) ($T_R\sim29.7$~K and $\sim30.6$~K respectively).  Under these conditions, the CO abundance traced is less than $3\times10^{-6}$.
We conclude that it is difficult for excitation alone to reconcile the emission with a CO abundance of 10$^{-4}$.

\section{$\chi$(CO) Measurement Caveats}
The analysis above assumes HD and CO emit from similar regions and therefore trace the gas-phase $\chi$(CO) directly. In the following section we relax this assumption and discuss various physical mechanisms that could modify the interpretation of the measured $\chi$(CO).
\label{caveats}
\subsection{Different Emitting Regions}\label{sec:der}
\begin{figure}
\epsscale{1.0}
\plotone{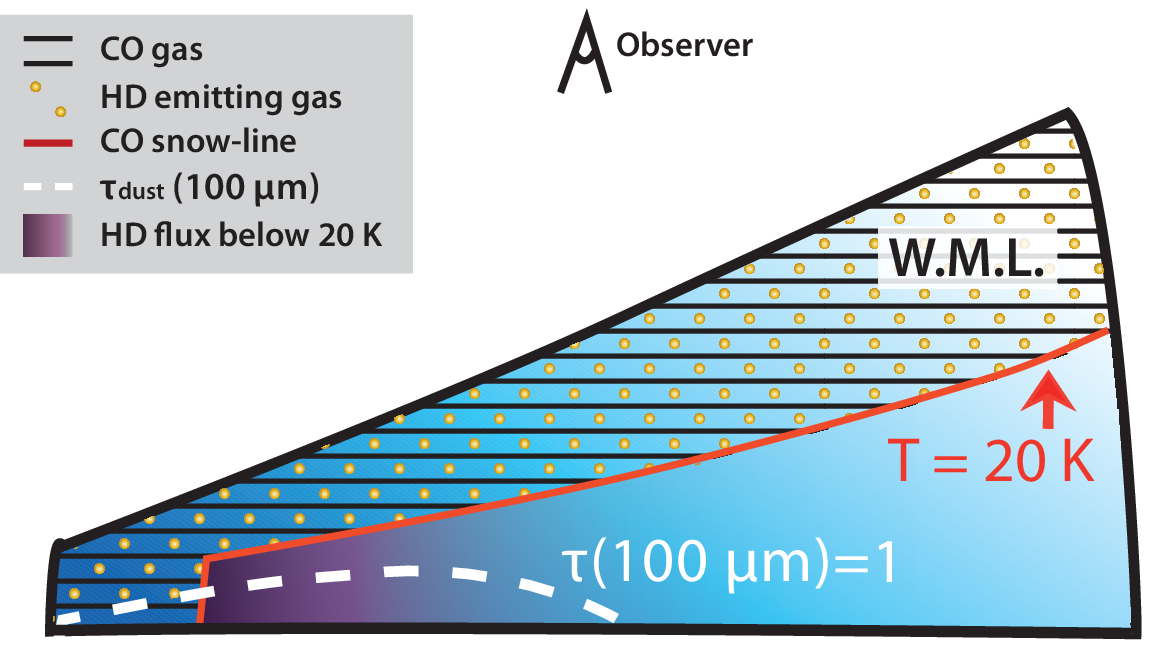}
\caption{Schematic illustrating the regions that contribute to HD and C$^{18}$O emission.  Horizontal black lines denote C$^{18}$O emitting region, $T_g>20$~K. We indicate the warm molecular layer (W.M.L.), the zone where CO is present in the gas. The yellow-dotted region denotes the HD~($1-0$) emitting region, generally restricted to $T_g>20$~K because of excitation considerations.  The magenta region denotes layers where HD could emit below 20~K, provided the midplane is sufficiently massive.  However, the midplane dust can become optically thick at 112~$\mu$m in a portion of this layer (denoted as the white dashed line), blocking HD emission from below.\label{fig4}}
\end{figure}

In Fig.~\ref{fig4} we illustrate some of the key issues concerning the above discussion.  First, while HD is spatially distributed broadly, gas-phase C$^{18}$O is not, freezing onto dust grains with $T_{\rm{dust}}<20$~K.  Because of the strong temperature dependence in the Boltzmann factor for the $J=1$ state, we would expect the HD emission to be sharply curtailed below $T_g\lesssim20$~K. For a massive midplane, some HD emission could arise from dense gas directly behind the CO snow-line (shown as magenta), but the HD emissivity from such cold gas is lessened by the fact that adding more mass (or enriching the dust) would increase the dust optical depth at 112~$\mu$m, hiding some fraction of the HD emission.  Furthermore, this emission cannot contribute significantly to the observations as it would drive the H$_2$ mass to unrealistically high levels.  For example, if $\sim20$\% of the HD~($1-0$) emission arises from gas at 15~K, the H$_2$ mass at this temperature is 0.05~M$_{\odot}$ in addition to the contribution from the rest of the disk.  Therefore it is difficult for the 15~K mass to add appreciably to the emission without driving the disk to extremely high masses.  

Another likely scenario is where the HD gas emits from primarily warm gas in the innermost disk, while CO and C$^{18}$O trace cooler emitting regions and thus larger physical radii. As a result, CO would trace more gas (full disk) than HD (warm inner disk). 
Consequently, HD~($1-0$) would miss H$_2$ mass in the outer disk, resulting in a lower limit to the disk mass estimation and in turn an overestimate of $\chi$(CO). The CO abundance could hence be lower. 
In addition, it is important to note that in B13 the authors find the outer disk does not emit appreciably, with only $\sim10\%$ of the HD flux coming from outside of 100~AU (see their Fig.~2c) based upon the model of \citet{Gorti:2011}.

\subsection{Freeze-out}\label{sec:freeze}
Previous studies have attributed measured low CO abundances to gas-phase depletion by adsorption onto grains \citep{Aikawa:1996,Dartois:2003}.  Under normal conditions CO freezes-out at low temperatures present in the midplane, $T\la20$~K, where HD does not strongly emit, and therefore the reduced measured $\chi$(CO) in the gas-phase is unlikely to be the result of freeze-out.

In fact a number of studies find the measured CO antenna temperatures of $T<17$~K \citep{Pietu:2007,Dartois:2003,Hersant:2009}.  If these estimates are correct, then the total volume of gas traced by the C$^{18}$O line exceeds that traced by the HD line, leading to an over-prediction of the true $\chi$(CO).  

There is, however, uncertainty in the freeze-out temperatures, which depend formally on the binding energies assumed.  The binding energies are a function of the binding-surface, often assumed to be CO ice. Alternatively, if the grain surface is water ice or bare dust, the binding energy can be significantly higher \citep{Bergin:1995,Fraser:2004}.  If this is the case, CO can freeze-out at higher temperatures $T>25$~K, and therefore the CO emitting region would be smaller than the HD emitting region.  In this instance the measured CO abundance would be lower than the true CO abundance.  

\subsection{Opacity}
Another caveat of our $\chi$(CO) estimates are the opacities of the HD~($1-0$) and C$^{18}$O~($2-1$) lines. In this study, we assume that emission of both species is optically thin.  Although we show in Sec.~\ref{sec:opa} that the C$^{18}$O~($2-1$) emission is thin in the disk-averaged data, the possibility of optically thick HD emission still remains. However, if $\tau_{\rm{HD}}\ga1$, the derived HD mass should be a lower limit and therefore the measured $\chi$(CO) is an upper limit on the true CO abundance. 

\subsection{Photodissociation and Self-shielding}
Photodissociation by UV is a major CO destruction mechanism in disks that regulates the molecular abundance of species in the gas. Photodissociation models for HD and CO isotopologues have been investigated by \citet{Roueff:1999,Le-Petit:2002,Visser:2009}.  \citet{Roueff:1999} finds HD should self-shield at smaller $A_V$ than CO.  Therefore, in the absence of dust shielding and selective isotopologue photodissociation, HD could emit from warm layers where C$^{18}$O is destroyed.  If those surface layers are essential contributors to the HD emission, $\chi$(CO) would be underestimated.  However, the modeling of B13 suggests that the high surface layers do not dominate the emissive mass of HD, and therefore, even if photodissociation cannot be ruled out, it only minimally affects the measured $\chi$(CO). Alternatively, if selective isotopologue photodissociation operates for C$^{18}$O from external UV irradiation, we may be missing CO mass from the outer disk edge.  As discussed in Section \ref{sec:der}, however, the outer disk does not significantly contribute to the HD emission.

%

\section{Implications: Where is the Carbon?}
\label{sec:implications}
Our study shows that the main reservoir of gas-phase carbon, CO, is reduced by at least an order of magnitude in the TW~Hya disk compared to dense clouds. In both T-Tauri and Herbig Ae disks similarly low CO abundances have been inferred and attributed to photodissociation and freeze-out \citep[e.g.,][]{Dutrey:2003,Chapillon:2008,Qi:2011}.
The difference between the previous studies and the results reported here is the use of HD to probe H$_2$ above 20~K and hence provide stronger constraints on $\chi$(CO) in the warm molecular layer. It is important to state that {\em{both}} C$^{18}$O and HD do not trace the midplane of the disk because of freeze-out (C$^{18}$O) and low excitation (HD). Thus it is possible that the $\chi$(CO$\rm_{ice}$) is ``normal'' in the midplane, which would be consistent with the similarity between interstellar ices and cometary volatiles \citep{mc11}. 
We argue differences in photodissociation of C$^{18}$O and HD are unlikely to account for the low $\chi$(CO). This would argue against the possibility that the carbon is sequestered in atomic form either neutral or ionized. \citet{Bruderer:2012} supports this assertion with observations of all primary forms of carbon in a Be star disk (HD~100547).  They argue the total carbon abundance is depleted in the warm atmosphere, which is consistent with our conclusion.

This finding leads one to ask where the missing carbon might be found. One possibility is suggested by the modeling of kinetic chemistry in disks by \citet{Aikawa:1997}. The deep disk layers are exposed to X-rays from the central star \citep{glassgold97}, though likely not cosmic rays \citep{cleeves13}. In these layers CO can exist in the gas via thermal- or photo-desorption from grains. X-rays produce He$^+$ and, with sufficient time, carbon can be extracted from CO via reactions with He$^+$.  CO reforms, but a portion of the carbon is placed into hydrocarbons (C$_{\rm{X}}$H$_{\rm{X}}$) or CO$_2$.  Many of these species have freeze-out temperatures higher than CO and trap the carbon in ices.  In a sense the chemistry works towards the first carbon-bearing molecule that freezes-out, creating a carbon sink \citep{Aikawa:1997}. Therefore we suggest that the low measured gas-phase CO abundance in the TW~Hya disk is a result of this chemical mechanism, and the use of CO as a mass tracer has very significant, and likely time-dependent, uncertainty.

\acknowledgments
We thank the anonymous referee for raising interesting issues. This work was supported by the National Science Foundation under grant 1008800. This paper makes use of the following ALMA data: ADS/JAO.ALMA2011.0.00001.SV and SMA data. ALMA is a partnership of ESO (representing its member states), NSF (USA) and NINS (Japan), together with NRC (Canada) and NSC and ASIAA (Taiwan), in cooperation with the Republic of Chile. The Joint ALMA Observatory is operated by ESO, AUI/NRAO and NAOJ. The Submillimeter Array is a joint project between the Smithsonian Astrophysical Observatory and the Academia Sinica Institute of Astronomy and Astrophysics and is funded by the Smithsonian Institution and the Academia Sinica.

{\it Facilities:} \facility{SMA}, \facility{Herschel Space Observatory Photodetector Array Camera and Spectrometer}, \facility{ALMA}


\end{document}